\documentstyle[aps,twocolumn]{revtex}
\begin{document}
\draft
\def\be{\begin{equation}}
\def\ee{\end{equation}}
\def\ba{\begin{eqnarray}}
\def\ea{\end{eqnarray}}
\title{Phase shift operator and cyclic evolution in finite dimensional
Hilbert space} 
\author{Ramandeep S. Johal\thanks{e-mail: 
raman\%phys@puniv.chd.nic.in}}  
\address{{\it Department of Physics, Panjab University,}\\ 
{\it Chandigarh -160 014, India. }} 
\date{\today}
\maketitle
\begin{abstract}
We address the problem of phase shift operator acting as
time evolution operator in Pegg-Barnett formalism. It is 
argued that standard shift operator is inconsistent 
with the behaviour of the state vector under cyclic
evolution. We consider  
 a generally deformed oscillator algebra at $q$ root
of unity, as it  yields the same Pegg-Barnett phase 
operator and show that shift operator within this
algebra meets our requirement.     
\end{abstract}
In recent years, Pegg-Barnett (PB) formalism has attracted wide
attention as a theory for quantum phase \cite{pb1},\cite{pb2}.
  Alongside, the subject
of quantum  algebras and their realizations in terms of
$q$-deformed oscillators has also been studied with great interest
\cite{qalg1}- \cite{qalg3}.
The problem of quantum phase has also been persued from
$q$-deformation theoretic point of view. There are certain
justifications for this approach. A feature of a $q$-deformed
theory or framework is that one can identify an inherent
scale in it, of magnitude $\sim |1-q|$, where $q$ is called
deformation parameter. As $q\to 1$, one retreives the 
undeformed or "classical" theory. Now PB formalism can 
also be looked upon as inherently $q$-deformed  \cite{abe1} in the
abovesaid sense and in this case $q={\rm exp}(i2\pi/s+1)$,
where $(s+1)$ is the dimension of the Hilbert space. 
Secondly, the phase observable which is hermitian phase
operator in PB theory, can be consistently defined only in a 
finite dimensional Hilbert space (FDHS). These two 
features form the motivation to study phase
using a $q$-oscillator with $q={\rm exp}(i2\pi/(s+1))$ \cite{elli}.
Firstly, $q$ being root of unity, naturally truncates
the $q$-oscillator to a FDHS. Secondly, infinite
dimensional limit ($s\to \infty$) also corresponds to 
the deformation free ($q\to 1$) limit. However, the problem
of negative norm in this representation was 
recognised later and there now exist representations
of $q$-oscillator \cite{fuji1} or generally deformed oscillator \cite{fu}
with positive norm  for $q$ as root of unity and for
which the Pegg-Barnett phase operator can be consistently
defined.   

Let us first recapitulate relevant key points of PB formalism.
Here the phase operator $\Phi$ and the number operator
$N$ are not canonically conjugate, but satisfy a complicated
commutator
\be
[{\Phi},N] = \frac{2\pi \hbar}{s+1}
\sum_{n,n^{\prime} =-l
}^{l}\frac{(n^{\prime}-n)|n^{\prime}\rangle\langle n|}
{{\rm exp}[2\pi i(n-n^{\prime})/(s+1)]-1}.
\label{com}
\ee
The eigenstates of $\Phi$ which form  an orthonormal
set of phase states, are related to the number states
by Fourier transform
\be
|\theta_m\rangle = \frac{1}{\sqrt{s+1}} \sum_{n=0}^{s}
{\rm exp}(in\theta_m)|n \rangle,
\label{ft}
\ee
where $\theta_m = \theta_0 + \frac{2\pi m}{s+1}$,
$\{m=0,1,2,...,s\}$.
$\theta_0$ is the arbitrary phase window which defines the
phase angle interval $2\pi$ modulo, $\theta_0\le \theta_m 
<\theta_0 + 2\pi$. 
Apart from the hermitian phase operator $\Phi$, the unitary
phase operator $e^{i{\Phi}}$ is also of significance in
PB theory. It acts as shift operator on number states
\ba
e^{i{\Phi}}|n\rangle  &=& |n-1\rangle, \qquad n \ne 0 \\
e^{i{\Phi}}|0\rangle  &=& e^{i(s+1)\theta_0}|s\rangle. \label{p2}
\ea
Thus the action of $e^{i{\Phi}}$ is cyclic and it steps
down the number states by unity. Its adjoint acts as step
up operator. Thus one can write a realization of unitary
phase operator as
 \be
e^{i\Phi} = |0 \rangle \langle 1| + |1 \rangle \langle 2| +
             \cdots + |s-1 \rangle \langle s| +
             e^{i(s+1)\theta_0} |s \rangle \langle 0|.
\ee
The operator dual to $e^{i{\Phi}}$ is the operator $q^N$, which
acts as shift operator on the phase states
\ba
q^{-N}|\theta_m\rangle &=& |\theta_{m-1}\rangle, \quad m\ne 0 \label{n1} \\
q^{-N}|\theta_0\rangle &=& |\theta_{s}\rangle. \label{n2}
\ea
Note that the apparent duality between the two kinds of shift
operators seems incomplete due to the extra phase factor in
Eq. (\ref{p2}) or the lack of corresponding factor in Eq. (\ref{n2}). 
This is due to the arbitrariness in the choice of phase window in
PB formalism, while there is no such choice in the ground state 
eigenvalue of number operator, which is necessarily zero.
Thus the realization of $q^{-N}$ in terms of phase states is
 \be
q^{-N} = |\theta_{0} \rangle \langle \theta_{1} | + |\theta_{1} \rangle 
             \langle \theta_{2} | +
             \cdots + |\theta_{s-1 } \rangle \langle \theta_{s} | +
              |\theta_{s} \rangle \langle \theta_{0} |.
\label{nrep}
\ee
Now the unitary phase shift operator $q^{-N}$ can be thought as 
time evolution operator, which operated once  on the phase state 
advances the phase
by $2\pi/(s+1)$. Thus if we operate it $(s+1)$ times on a phase
state, we complete one cycle and return to the same phase state.
On the other hand, we have the results of Ref. \cite{pati},
where it was shown that for cyclic evolution of harmonic
oscillator  in a general state $\sum_{n} c_n|n\rangle$, in FDHS,
the state vector can change sign which depends on the dimesnionality
of the space; if $(s+1)$ is even sign changes, otherwise not.
However, if we take $q^{-N}$ as equivalent to time evolution operator,
we note that according to realization of Eq. (\ref{nrep}), the 
state vector always returns exactly to initial state, irrespective
of the dimensionality of the space.     

The purpose of this paper is to make the action of 
phase shift operator consistent with that of
time evolution operator in the context of cyclic
evolution in FDHS. We take as our model the recently proposed generally
deformed oscillator \cite{fu}, which has certain advantages
over other approaches from algebraic point of view, namely, 
i) The creation and annihilation operators in PB theory
do not form a closed algebra by themselves, and they do
not go over to corresponding relations in the $s\to \infty$
limit ii) we can algebraically define PB phase operator 
in the approach of \cite{fu}, iii) for $q$ as root
of unity, positive norm is also assured.     
  
Briefly, in the approach of \cite{fu}, new creation and
annihilation operators are defined
\be
A^{\dag} = \sqrt{{\cal F}(q^{\cal N})}e^{-i\Phi},\quad
A = e^{i\Phi}\sqrt{{\cal F}(q^{\cal N})}, \quad q^{\cal N} = q^{N+\eta}.
\label{ca}
\ee
The action of these operators on generalized number states is
\ba
A^{\dag}|n+\eta\rangle &=& \sqrt{{\cal F}(q^{n+\eta +1})}|n+\eta
+1\rangle, \qquad n\ne s \\
A^{\dag}|s+\eta\rangle &=& e^{-i(s+1)\theta_0} \sqrt{{\cal F}(q^{n})}|\eta\rangle
\label{a1}  \\
~~~A|n +\eta\rangle &=& \sqrt{{\cal F}(q^{n+\eta})} |n+\eta-1\rangle,\qquad n \ne 0 \\
~~~A|\eta\rangle &=& \sqrt{{\cal F}(q^{\eta})}e^{i(s+1)\theta_0}  |s+\eta\rangle 
\label{a2}\\ 
q^{\cal N}|n+\eta\rangle &=& q^{n+\eta}|n+\eta\rangle.
\ea
The parameter $\eta$ is chosen such that i) the above defines
a cyclic representation, ii) the function ${\cal F}$ is hermitian and 
non-negative, iii) in $s\to \infty$ limit, $A^{\dag}$ and $A$ go over
to the creation and annihilation operators of the ordinary oscillator.
Also the condition for cyclic representation (${\cal F}(q^{\eta})\ne 0$)
in Eqs. (\ref{a1}) and (\ref{a2})) also ensures that one can
consistently define unitary phase operator by inverting $A^{\dag}$ and $A$
in Eq.  (\ref{ca}). Note that this approach exactly recovers the
PB phase operator.

However a significant fact that was missed in \cite{fu} is that
in the above representation,  $q^{\pm {\cal N}}$ can also act
as phase shift operator on the phase states. As one can easily see,
 its action gives
$q^{-\cal N}|\theta_m\rangle= q^{-\eta}|\theta_{m-1}\rangle$
and $q^{-\cal N}|\theta_0\rangle =q^{-\eta} |\theta_{s}\rangle$,
which is just same as Eqs. ({\ref{n1}}) and ({\ref{n2}}). 
However, as we show below, the significance of this operator 
 lies in its being  
consistent with the results of cyclic evolution in FDHS \cite{pati}.
 As a solution for
restoring the duality in $e^{i\Phi}$ and $q^{-\cal N}$, we 
propose to modify the Eq. (\ref{ft}) as follows:  
\be
|\theta_m\rangle = \frac{1}{\sqrt{s+1}} \sum_{n=0}^{s}
{\rm exp}(i(n+\eta)\theta_m)|n+\eta \rangle,
\label{ndef}
\ee
so that now we have   
\ba
q^{-\cal N}|\theta_m\rangle &=& |\theta_{m-1}\rangle, \quad m\ne 0 \\
q^{-\cal N}|\theta_0\rangle &=&   e^{-i2\pi\eta} |\theta_{s}\rangle.
\ea
The action of $e^{i\Phi}$ on the phase states  remains as such, i.e.
$e^{i\Phi}|\theta_m\rangle = \theta_m |\theta_m\rangle$.
Moreover, the action of  $e^{i\Phi}$ on the (new) number states
remains same as before. Thus from Eq. (\ref{ndef})
\be
|n+\eta\rangle = \frac{1}{\sqrt{s+1}}\sum_{m=0}^{s} {\rm exp}(-i(n+\eta)
\theta_m) |\theta_m\rangle,
\label{ndef2}
\ee
we can write
\ba
e^{i{\Phi}}|n+\eta\rangle  &=& |n+\eta-1\rangle, \qquad n \ne 0 \\
e^{i{\Phi}}|\eta\rangle  &=& e^{i(s+1)\theta_0}|s+\eta\rangle. 
\ea
Thus the  duality between $e^{i\Phi}$ and $q^{-\cal N}$
is exactly obeyed, so that parameter $\eta$ plays the role 
equivalent to $\theta_0$. We can as well write the
following realization for modified unitary operator
 \be
q^{-\cal N} = |\theta_{0} \rangle \langle \theta_{1} | + |\theta_{1} \rangle
             \langle \theta_{2} | +
             \cdots + |\theta_{s-1 } \rangle \langle \theta_{s} | +
              e^{-i2\pi\eta}  |\theta_{s} \rangle \langle \theta_{0} |.
\ee
Therefore, operating the above unitary operator $(s+1)$ times, we get 
\be
\left(q^{-\cal N}\right)^{s+1}|\theta_m\rangle=
     e^{-i2\pi\eta} |\theta_m\rangle.
\ee
Next, we are interested to know if under such cyclic evolution, the 
state vector changes sign or not. Thus if $\eta$ is an integer,
no change in sign occurs, while for $\eta$ as half-odd integer, 
there is change in sign. 
Now the  usual time evolution operator
is $e^{-iHt/\hbar}$, where for the case of harmonic oscillator in FDHS \cite{pati},
the hamiltonian $H$ has the following energy spectrum
\be
E_n = \hbar\omega\left( n +\frac{1}{2} + \frac{(s+1)}{2}\delta_{n,s}\right).
\ee
Thus under evolution through one time period, $t = 2\pi/\omega$,
the state vector $|n\rangle$ is multiplied by the phase 
factor       
${\rm exp}\;(-i2\pi\{n+1/2+(s+1)\delta_{n,s}/2\})$.
On the other hand, if we consider time evolution through
unitary shift operator $q^{-{\cal N}}$, this means that
state vector is multiplied by the factor  
${\rm exp}\;(-i2\pi\{n+\eta\})$. 
Thus we note that for a harmonic oscillator in FDHS, for
$n\ne s$, we have $\eta =1/2$, whereas for $n=s, \eta = 1/2+(s+1)/2$.
So $(s+1)$ as even number is equivalent to $\eta$
as half-odd integer, which from previous discussion,
  implies change in sign under
cyclic evolution, whereas $(s+1)$ odd is equivalent
to $\eta$ as integer and consequently no change in sign
of the state vector under one cycle.
Also, the case of infinite dimensional harmonic oscillator
requires that $E_n = (n+1/2)\hbar\omega$, which is
consistent with $\eta = 1/2$.

Finally, it is interesting to note that states $|n+\eta\rangle$
can be obtained from usual number states $|n\rangle$,
by applying a continuous unitary transformation i.e.
when $\eta$ is not an integer
 ($ e^{-i\eta\Phi}|n\rangle = |n+\eta\rangle$).
As was pointed out in \cite{pb1}, such continuous  unitary transformations
are useful to construct the  phase-moment generating functions. 

Concluding, we have argued that phase shift operator in standard
PB formalism is inconsistent with cyclic evolution
of harmonic oscillator in finite dimesional Hilbert 
space. To treat this, we have shown that phase
shift operator of a generally deformed oscillator
algebra at $q$ root of unity, and which yields
the same PB phase operator, can simulate 
the behaviour of time evolution operator  for
cyclic evolution. This also restores the duality in the
actions of phase- and number- shift operators.  

The author would like to acknowledge the kind hospitality
of H.S. Mani and Sumathi Rao at Mehta Research Institute, Allahabad,
where this work was initiated
and S. Abe, for careful reading of the manuscript.

\end{document}